\documentclass[10pt,conference]{IEEEtran}
\usepackage{blindtext,graphicx}
\usepackage{times}
\usepackage[a4paper, total={184mm,239mm}]{geometry}
\IEEEoverridecommandlockouts
\usepackage{cite}
\usepackage{amsmath,amssymb,amsfonts,bm}
\usepackage[noend]{algorithmic}
\usepackage{textcomp}
\usepackage{flushend}
\usepackage{setspace}
\def\smallerspacecaption{\vspace{-2mm}}

\usepackage{floatrow}
\usepackage[caption=false]{subfig}
\usepackage[table, dvipsnames]{xcolor}  
\usepackage{fancyhdr}
\usepackage{graphicx}
\usepackage{url}
\usepackage{verbatim}
\usepackage{multirow}
\usepackage{hhline}
\usepackage[us]{datetime}
\usepackage{paralist}
\usepackage{color}
\usepackage{soul}
\usepackage{enumitem}
\usepackage[implicit=false,hidelinks]{hyperref}
\usepackage{stfloats}
\usepackage[font=footnotesize,labelsep=period,justification=centering]{caption}
\usepackage[ruled, vlined, norelsize]{algorithm2e} 
\usepackage{pifont}
\graphicspath{{incl/}}

\newcommand{\blue}[1]{\textcolor{black}{#1}}

\usepackage{soul}
\newcommand{\drop}[1]{\textcolor{red}{#1}}
\renewcommand{\drop}[1]{}
\definecolor{cadmiumgreen}{rgb}{0.0, 0.42, 0.24}
\newdimen\arrayruleHwidth
\makeatletter
\def\Hline{\noalign{\ifnum0=`}\fi\hrule \@height \arrayruleHwidth
\futurelet \@tempa\@xhline}
\makeatother

\makeatletter
\def\blfootnote{\xdef\@thefnmark{}\@footnotetext}
\makeatother

\shortdate
\settimeformat{ampmtime}

\makeatletter
\newcommand*\tabsize{%
	   \@setfontsize\tabsize{6}{7.2}%
}
\makeatother

\renewcommand{\IEEEbibitemsep}{0.3pt}
\makeatletter
\IEEEtriggercmd{\reset@font\normalfont\fontsize{6.8pt}{7.11pt}\selectfont}
\makeatother
\IEEEtriggeratref{1}

\makeatletter
\renewcommand\footnoterule{%
  \kern-3\p@
  \hrule\@width 0.5\columnwidth
  \kern2.6\p@}
  \makeatother

\usepackage{amsmath,amsfonts,bm}

\newcommand{\G}[1][]{
    \IfEq{#1}{}
    {\mathcal{G}}
    {\mathcal{G}}_{#1}}
\newcommand{\V}[1][]{
    \IfEq{#1}{}
    {\mathcal{V}}
    {\mathcal{V}_{#1}}}
\newcommand{\E}[1][]{
    \IfEq{#1}{}
    {\mathcal{E}}
    {\mathcal{E}_{#1}}}
\newcommand{\N}[1][]{
    \IfEq{#1}{}
    {\mathcal{N}}
    {\mathcal{N}}_{#1}}
\newcommand{\X}[1][]{
    \IfEq{#1}{}
    {\bm{X}}
    {\bm{X}^{\paren{#1}}}}
\newcommand{\Asym}[1][]{
    \IfEq{#1}{}
    {\widetilde{\bm{A}}}
    {\widetilde{\bm{A}}_{#1}}}
\newcommand{\Arw}[1][]{
    \IfEq{#1}{}
    {\widehat{\bm{A}}}
    {\widehat{\bm{A}}_{#1}}}
\newcommand{\A}[1][]{
    \IfEq{#1}{}
    {\bm{A}}
    {\bm{A}}_{#1}}
\newcommand{\Hx}[1][]{
    \IfEq{#1}{}
    {\bm{H}}
    {\bm{H}}^{(#1)}}

\newcommand{\W}[1][]{
    \IfEq{#1}{}
    {\bm{W}}
    {\bm{W}^{\paren{#1}}}}

\newcommand{\paren}[1]{\left( #1 \right)}

\def\mA{{\bm{A}}}
\def\mB{{\bm{B}}}

\def\mD{{\bm{D}}}

\def\mH{{\bm{H}}}
\def\mI{{\bm{I}}}

\def\mX{{\bm{X}}}

\DeclareMathAlphabet{\mathsfit}{\encodingdefault}{\sfdefault}{m}{sl}
\SetMathAlphabet{\mathsfit}{bold}{\encodingdefault}{\sfdefault}{bx}{n}

\begin{document}

\title{\huge MuxLink: Circumventing Learning-Resilient MUX-Locking Using Graph Neural Network-based Link Prediction \vspace{-5pt}}

\author{Lilas~Alrahis$^\ddag$, Satwik~Patnaik$^\dag$, Muhammad~Shafique$^\ddag$, and Ozgur~Sinanoglu$^\ddag$\\[1ex]
$^\ddag$Division of Engineering, New York University Abu Dhabi, UAE\\
$^\dag$Electrical \& Computer Engineering, Texas A\&M University, College Station, Texas, USA\\
\normalsize{\{lma387, muhammad.shafique, ozgursin\}@nyu.edu}, \normalsize{satwik.patnaik@tamu.edu}
}
\maketitle
\renewcommand{\headrulewidth}{0.0pt}
\thispagestyle{fancy}
\lhead{}
\rhead{}
\chead{\copyright~2022 IEEE.
This is the author's version of the work.
The definitive Version of Record will be published in Proc. Design, Automation and Test in Europe (DATE) 2022}
\cfoot{}
\begin{abstract}
Logic locking has received considerable interest as a prominent technique for protecting the design intellectual property from untrusted entities, especially the foundry. 
Recently, machine learning (ML)-based attacks have questioned the security guarantees of logic locking, and have demonstrated considerable success in deciphering the secret key without relying on an oracle, hence, proving to be very useful for an adversary in the fab. 
Such ML-based attacks have triggered the development of learning-resilient locking techniques. 
The most advanced state-of-the-art deceptive MUX-based locking (D-MUX) and the symmetric MUX-based locking techniques have recently demonstrated resilience against existing ML-based attacks. 
Both defense techniques obfuscate the design by inserting key-controlled MUX logic, ensuring that all the secret inputs to the MUXes are equiprobable.

In this work, we show that these techniques primarily introduce local and limited changes to the circuit without altering the global structure of the design. 
By leveraging this observation, we propose a novel graph neural network (GNN)-based link prediction attack, \textit{MuxLink}, that successfully breaks both the D-MUX and symmetric MUX-locking techniques, relying only on the underlying structure of the locked design, i.e., in an oracle-less setting.
Our trained GNN model learns the structure of the given circuit and the composition of gates around the non-obfuscated wires, thereby generating meaningful link embeddings that help decipher the secret inputs to the MUXes.
The proposed MuxLink achieves \blue{key prediction} accuracy and precision up to $100\%$ on D-MUX and symmetric MUX-locked ISCAS-85 and ITC-99 benchmarks, \blue{fully unlocking the designs.} We open-source MuxLink~\cite{webinterface}.

\end{abstract}


\section{Introduction}
\label{sec:Introduction}

\begin{figure}[!t]
\centering
\includegraphics[width=\textwidth]{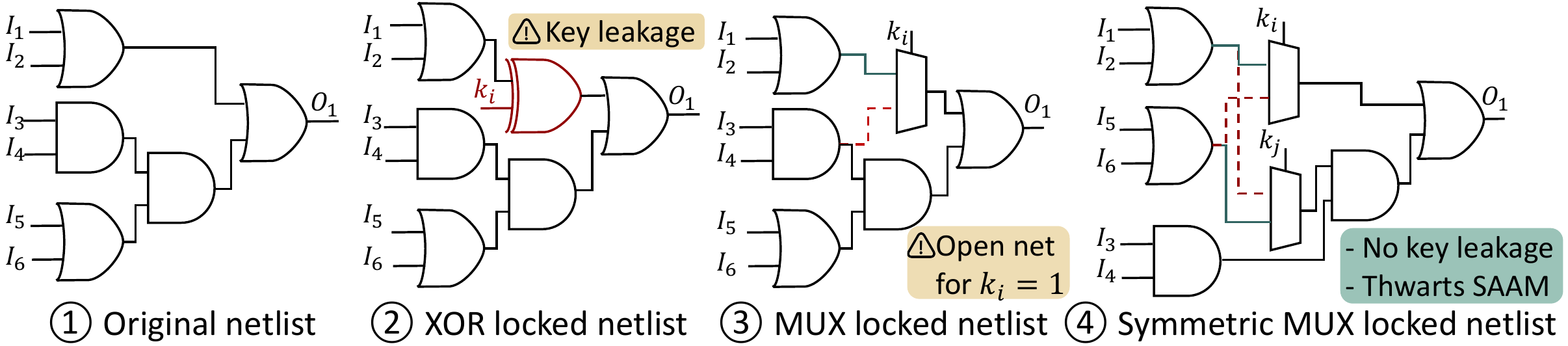}
\smallerspacecaption
\vspace{-0.6em}
\caption{Examples of XOR logic locking (LL) and MUX-based LL.}
\vspace{-0.6em}
\label{fig:example}
\end{figure}

The globalized and, thus, distributed semiconductor supply chain creates an attack vector for the untrusted entities in stealing the intellectual property (IP) of a design. 
To ward off the threat of IP theft, researchers developed various countermeasures like state-space obfuscation, split manufacturing, and logic locking (LL). 
LL entails inserting additional key logic (XOR/XNOR gates, MUXes) in the original design.
These added logic gates \blue{(referred to as key-gates)} are driven by a secret key (known only to the designer) through an on-chip tamper-proof memory. 
We illustrate examples of LL in Fig.~\ref{fig:example}. 
The prime reason behind the widespread prevalence of LL is because it protects the design IP throughout the supply chain (foundry, test facility, end-users).

Researchers have developed various attacks on LL to recover the secret key, considering two threat models: the \textit{oracle-guided} and the \textit{oracle-less}. 
The oracle-guided attacks require a functional chip (with the secret key embedded) acting as an \textit{oracle}~\cite{Subramanyan_host_2015}, which may not be available in many practical settings. Towards a more realistic scenario, the oracle-less attacks rely only on the structure of the locked design, posing a significant threat to LL. 
Recently, machine learning (ML) algorithms have facilitated various oracle-less attacks on LL~\cite{sail,alrahis2021unsail,snapshot,gnnunlock_tetc,OMLA,alrahis2021untangle} leading to the development of learning-resilient LL to counteract these threats~\cite{limaye2020thwarting,alrahis2021unsail,sisejkovic2021deceptive}. Other ML-based attacks~\cite{genunlock,bocanet,nngsat} follow the oracle-guided model. We focus on the oracle-less threat model, which is a more challenging attack model.

\begin{figure}[!t]
\centering
\includegraphics[width=\textwidth]{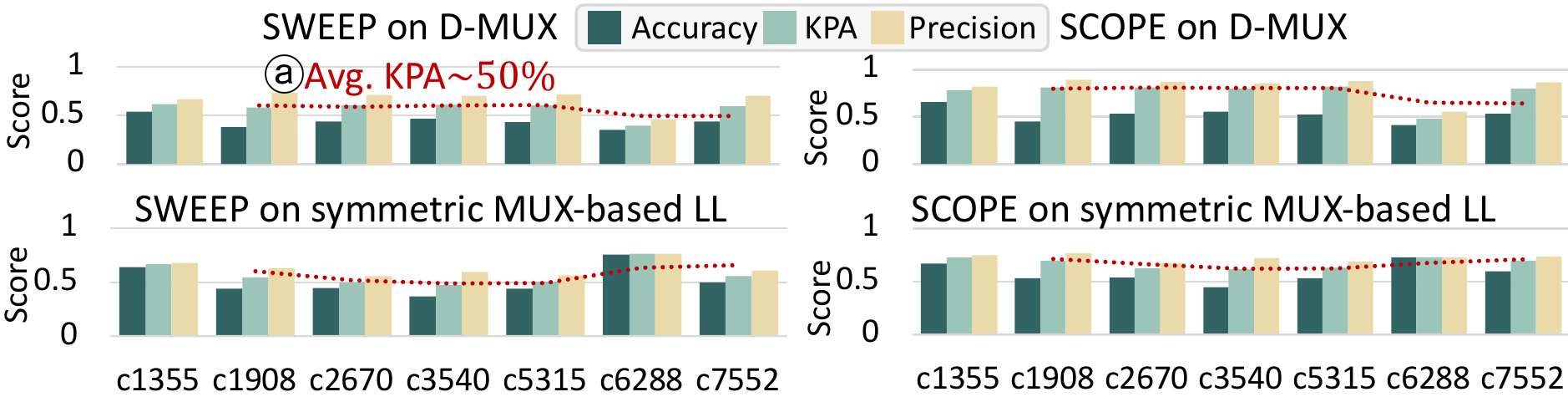}
\smallerspacecaption
\vspace{-0.6em}
\caption{
D-MUX~\cite{sisejkovic2021deceptive} and symmetric MUX-based LL~\cite{alaql2021scope} 
are resilient to the constant propagation attacks SWEEP~\cite{sweep} and SCOPE~\cite{alaql2021scope}.}
\vspace{-0.5em}
\label{fig:motivation}
\end{figure}

\textbf{In this work}, we focus on the deceptive MUX-based (D-MUX)~\cite{sisejkovic2021deceptive} and the symmetric MUX-based LL techniques~\cite{alaql2021scope}. In the following, first, we discuss the aspects of D-MUX and symmetric MUX-based locking which make them learning-resilient and outline the key research challenges for our work.

\subsection{Key Research Challenges Targeted in this Work}

\begin{enumerate}[leftmargin=*]

\item \textbf{No key leakage:} Existing ML-based attacks~\cite{sail,snapshot,OMLA} capture the local locking-induced structural modifications, which embed key information to decipher the secret key. 
An example of XOR LL is depicted in Fig.~\ref{fig:example}\raisebox{.5pt}{\textcircled{\raisebox{-.9pt}{2}}}, where there is a direct mapping between the type of the key-gate and the key-value. Although re-synthesis is performed to induce local transformations around the key-gate, thereby breaking the mapping between the type of the key-gate and key-value, potent ML-based attacks such as \textit{SAIL}~\cite{sail}, \textit{SnapShot}~\cite{snapshot} and \textit{OMLA}~\cite{OMLA} expose the key information from the surrounding circuitry.
Learning-resilient MUX-based locking eliminates the key-related leakage in the structural residue from locking, i.e., the location selection of the key-logic and the introduced changes, thwarting existing ML-based attacks. 
The authors in~\cite{sisejkovic2021deceptive} launched SnapShot~\cite{snapshot} on D-MUX locked benchmarks and demonstrated that the attack reports a consistent key prediction accuracy (KPA) around $50\%$, which means that D-MUX forces SnapShot to perform random guesses about the 
secret key.

\item \textbf{No circuit reduction:} A MUX key-gate takes in two wires from the design as inputs. 
A key-input acts as the select line, passing the true wire upon applying the correct key-value. 
One major vulnerability in na\"{\i}ve MUX-based locking is that a wrong key-bit could remove an entire logic cone (the true cone) from the circuit.
An example is illustrated in Fig.~\ref{fig:example}\raisebox{.5pt}{\textcircled{\raisebox{-.9pt}{3}}}, where setting $k_i=1$ disconnects the true wire (green). 
The structural analysis attack \textit{SAAM}~\cite{sisejkovic2021deceptive} exposed this vulnerability. 
The learning-resilient MUX-based locking techniques ensure no reduction in the design for the wrong key-values and are completely resilient to SAAM.

\item \textbf{Symmetric paths:} Na\"{\i}ve MUX-based locking is vulnerable to the constant propagation attacks \textit{SWEEP}~\cite{sweep} and \textit{SCOPE}~\cite{alaql2021scope}. 
The attacks hard-code the value of one key-bit at a time and perform re-synthesis. 
A difference in the design features, including power consumption, total area, etc., is observed in the re-synthesized circuits, which correspond to the two possible values for a single key-input. 
Thus, the attacks learn the correlation between the extracted features and the correct key.
The D-MUX and the symmetric MUX-based LL add pairs of MUX key-gate, where the true logic cones of the MUXes are symmetrically interconnected, eliminating this kind of attack. 
In addition, ensuring no circuit reduction for wrong key-values (see the point above) enhances the resilience against constant propagation attacks. 
An example of symmetric MUX-based LL is shown in Fig.~\ref{fig:example}\raisebox{.5pt}{\textcircled{\raisebox{-.9pt}{4}}}.
We locked seven ISCAS-85 benchmarks using D-MUX and the symmetric MUX-based LL. We performed the same evaluation as in~\cite{sisejkovic2021deceptive}, where each circuit is copied $100$ times and locked with a key-size of $K=64$; resulting in $700$ locked benchmarks. 
The locked benchmarks are directly attacked using SCOPE as it does not require training. 
For SWEEP, we generate one dataset for each target benchmark, where the $100$ locked versions are kept for testing, while the $600$ other benchmarks and used for training. 
We repeat this analysis for both techniques, and the average accuracy, precision, and KPA for each benchmark are shown in Fig.~\ref{fig:motivation}.
The attacks report an average $KPA\approx50\%$ (see Fig.~\ref{fig:motivation}\raisebox{.5pt}{\textcircled{\raisebox{-.9pt}{a}}})\footnote{More details on the evaluation metrics are given in Sec.~\ref{sec:results}.} confirming the resilience of the techniques against existing learning-based and constant propagation attacks.
\end{enumerate}

\begin{figure}[!t]
\centering
\includegraphics[width=0.8\textwidth]{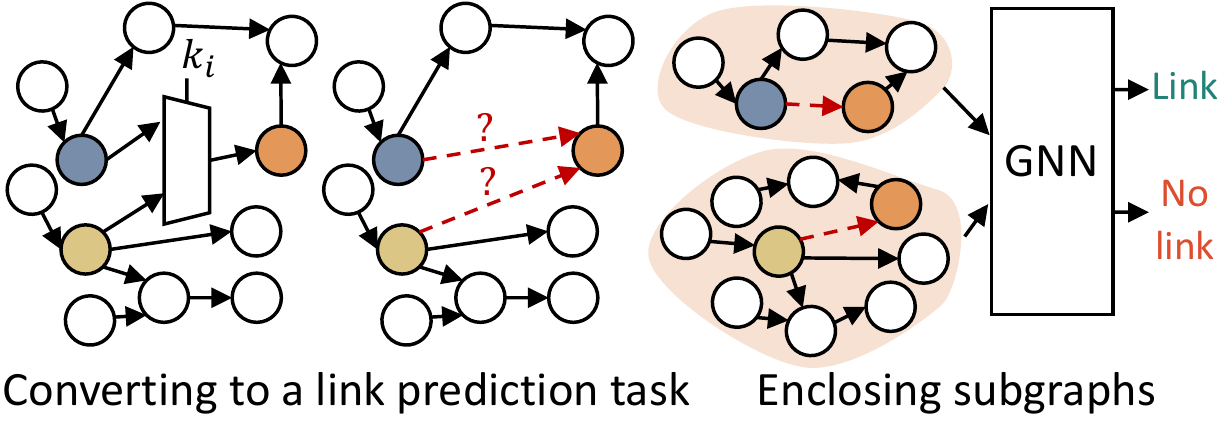}
\smallerspacecaption
\caption{Unlocking MUX-based obfuscation using GNN-based link prediction.}
\vspace{-0.4em}
\label{fig:idea}
\end{figure}

\subsection{Our Novel Research Contributions}

In this work, we propose the \textit{MuxLink} platform, that leverages efficient graph neural network (GNN)-based link prediction to devise the first-ever attack on the 
MUX-based learning-resilient LL.\footnote{\textit{GNNUnlock}~\cite{alrahis2021gnnunlock,gnnunlock_tetc} is a GNN-based oracle-less attack targeting SAT-resilient LL, in which the protection logic has a structure that can be learned by a GNN, isolated, and detached. D-MUX and the symmetric MUX-based LL do not have isolable logic, and thus, are not applicable to
GNNUnlock.}
We decipher the secret MUX inputs in an \textit{oracle-less} setting by exploiting the underlying structure of the target design \textit{without requiring any circuit library or re-locking for training}. 
We show that D-MUX and symmetric-MUX-based locking 
introduce only limited local modifications to the designs, and provide resilience against locality-based learning attacks specifically, but not \textit{necessarily} against \textit{any} learning-based approach. Our novel contributions are as follows.

\begin{enumerate}[leftmargin=*]

\item A \textbf{link prediction-based platform} is developed, in which the task of deciphering the secret MUX inputs is converted to a link prediction problem. We employ a GNN to learn meaningful link representations and decipher the true connections, recovering the original design.

\item A \textbf{key recovery post-processing} guided by the likelihood scores of the GNN is developed to recover the secret key.

\end{enumerate}

\textbf{Major Results:} We evaluate the effectiveness of MuxLink through an extensive experimental analysis on selected ISCAS-85 and ITC-99 benchmarks locked using D-MUX~\cite{sisejkovic2021deceptive} and symmetric MUX-based locking~\cite{alaql2021scope}. 
MuxLink deciphers up to $100\%$ of the key-bits with a precision up to $100\%$ \blue{in seconds}, unlocking benchmarks which the other state-of-the-art oracle-less attacks fail to unlock.
\textbf{We open-source MuxLink~\cite{webinterface}.}

\section{Background and Related works}
\label{sec:backgroud_RW}
\vspace{-0.2em}
\subsection{Tests for Learning-Resilent LL}

The authors in~\cite{sisejkovic2021deceptive} consider two types of designs to evaluate LL for learning resilience. 
The first category includes designs that are synthesized with only a single type of gate (e.g., AND gate). 
The second category includes synthesized designs constituting randomly selected and well-distributed logic gates. 
In addition, the authors propose two learning-resilience tests, AND netlist test (ANT) and random netlist test (RNT).
Suppose a locking technique fails either of the two tests. 
In that case, the locking technique in question is regarded as conclusively vulnerable as the resilience of the technique is governed by the secret key and the structure of the design.
\vspace{-0.2em}
\subsection{Initial Learning-Resilient LL Techniques}

UNSAIL~\cite{alrahis2021unsail} injects identical key-gate structures with differing key-bit values in the locked design, which leads to flawed inferences from the ML models used in the SAIL attack~\cite{sail}.
The key-gate insertion phase in UNSAIL is guided by a targeted re-synthesis procedure. A learning-resilient LL should ideally deliver security without having any dependence on the synthesis tool to obfuscate the locking-induced transformations~\cite{sisejkovic2021deceptive}.
\begin{figure}[!t]
\centering
\includegraphics[width=\textwidth]{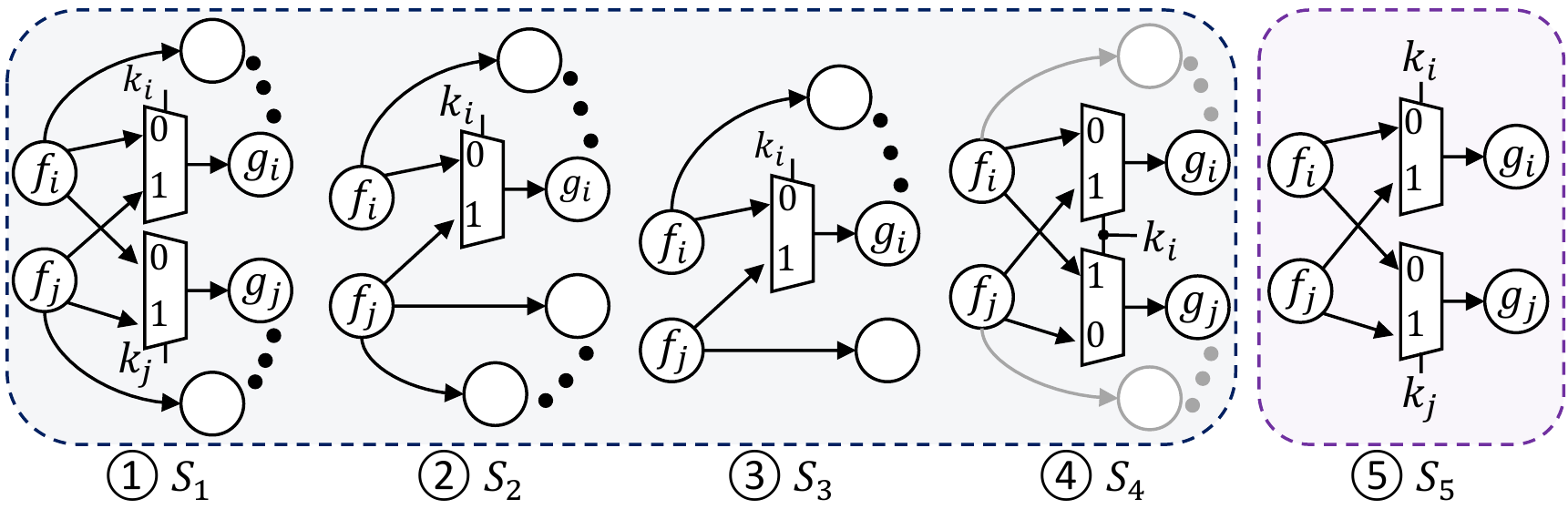}
\smallerspacecaption
\caption{$S_1\rightarrow S_4$ are the D-MUX locking strategies. $S_5$ is the symmetric MUX-based locking strategy.}
\vspace{-0.2em}
\label{fig:schemes}
\end{figure}

In~\cite{limaye2020thwarting}, the authors introduced truly random LL (TRLL), in which random decisions are made regarding the insertion of the key-gates. TRLL involves (i) replacement of inverters with an XOR key-gate, (ii) insertion of XOR key-gate at any location in the design, and (iii) insertion of XOR key-gate followed by an inverter. Although TRLL does not rely on synthesis tools and passes RNT, it fails ANT since there are no inverters to be replaced or coupled with an XOR gate, and this technique reduces to a conventional XOR-based LL technique.

\subsubsection{Deceptive MUX-based LL (D-MUX)~\cite{sisejkovic2021deceptive}}

Supporting both the ANT and RNT concepts remains a challenge for X(N)OR-based LL. The authors in~\cite{sisejkovic2021deceptive} state that vulnerabilities manifest due to the insertion of additional logic without leaving key-related, structural traces. Furthermore, the authors conjecture that MUX-based LL has an important advantage, as it inserts the same structure (i.e., a MUX), and it reconfigures the existing logic.
A new learning-resilient LL, D-MUX, 
is proposed~\cite{sisejkovic2021deceptive}, which ensures that each wire feeding to the MUX has the same probability of being true/false. Multiple locking strategies ($S_{1}$--$S_{4}$ in Fig.~\ref{fig:schemes}) are followed by D-MUX.

In the $S_{1}$ strategy, two multi-output nodes $\{f_{i}, f_{j}\}$ are selected as inputs to two locking MUXes. 
The MUXes obfuscate one randomly selected output node for each input node, i.e., $\{g_{i}, g_{j}\}$. 
Two individual key-inputs $\{k_{i}, k_{j}\}$ are used, where each key-input acts as a select line for one MUX. The $S_2$ strategy selects two multi-output nodes $\{f_{i}, f_{j}\}$, but performs locking using a single key-input $k_{i}$ controlling a single MUX. 
One randomly selected output node for a randomly selected input node is locked.
E.g., in Fig.~\ref{fig:schemes}\raisebox{.5pt}{\textcircled{\raisebox{-.9pt}{2}}}, $S_{2}$ selects $f_{i}$ and one of its output nodes $g_{i}$. 
The $S_3$ strategy selects and locks one multi-output node ${f_{i}}$ using a single key-input $k_{i}$ controlling one MUX. 
$f_j$ in the case of $S_3$ is a single-output node. 
Finally, the $S_{4}$ strategy sets no restrictions on $\{f_{i}, f_{j}\}$. 
A single key-input $k_{i}$ drives two MUXes and locks one output node for each input node.
In all the strategies, the MUXes are configured to cause no circuit reduction and no combinational loops.

The cost of the $S_4$ strategy, in terms of the number of gates added, is larger compared to the rest of the strategies. However, $S_4$ is always applicable as there are no restrictions on $\{f_i,f_j\}$. To reduce costs, the enhanced D-MUX (eD-MUX) only uses $S_4$ when none of the other strategies is viable.

\begin{figure*}[ht]
\centering
\includegraphics[width=0.9\textwidth]{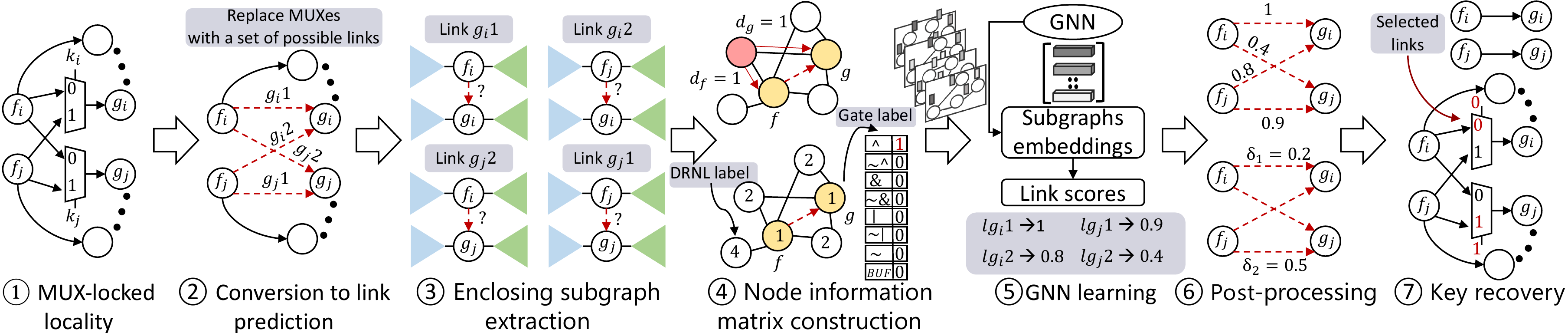}
\smallerspacecaption
\caption{The methodology work flow. 
The blue and green triangles represent fan-in cones and fan-out cones, respectively.}
\label{fig:MuxLink}
\end{figure*}

\subsubsection{Symmetric MUX-based LL~\cite{alaql2021scope}}
Concurrent to D-MUX, Alaql \textit{et al.}~\cite{alaql2021scope} propose another technique (Fig.~\ref{fig:schemes}\raisebox{.5pt}{\textcircled{\raisebox{-.9pt}{5}}}), which can be considered as a special case of D-MUX. 
We denote this locking strategy as $S_5$. 
Note that $S_{5}$ is equivalent to $S_{4}$, but two individual key-inputs are driving the individual MUXes. 
Here, $\{f_i, f_j\}$ are one-output nodes.

\vspace{-0.2em}
\subsection{Link Prediction Problem}

Link prediction refers to the problem of inferring missing links from an observed graph. 
Let $\mathcal{G}(\mathcal{D}, \mathcal{V})$ denotes a graph with a set of edges $\mathcal{D}$ and a set of nodes $\mathcal{V}$. 
Given $\mathcal{V}$ and a subset of true links $\mathcal{E}\in\mathcal{D}$, the objective is to identify the unobserved true links $\mathcal{S}$ referred to as \textit{target links}, where $\mathcal{D}=\mathcal{E}+\mathcal{S}$.
Link prediction has varied usages in recommender systems, drug discovery, and knowledge graph completion, etc.
Traditional link prediction heuristics rely on handcrafted features, which might fail to express the complex patterns in the graph that actually determine the link formations. 
Recently, the authors in ~\cite{SEAL} demonstrated how GNNs can 
directly learn suitable ``heuristics'' from local enclosing subgraphs around links. 
\textit{SEAL}~\cite{SEAL} extracts an enclosing subgraph around each target link, computes a subgraph embedding using a GNN, and uses it for link prediction.
Since the subgraph embeds information regarding the target link, the label of the subgraph can be considered the label of the target link.

\vspace{-0.2em}
\subsection{Graph Neural Networks (GNNs)}
\label{subsec:gnn}

GNNs generate an embedding for each node $v \in \mathcal{V}$ in a graph $\mathcal{G}$ through iterations of \textit{message passing}~\cite{zhang2018end} as follows, where $h_v^{l}$ denotes the embedding of $v$ at the $l^{th}$ iteration.
\begin{equation}
 \vspace{-0.2em}
 \footnotesize
 a_v^{l} = AGG^{l}(\{h_u^{l-1}: u \in N(v)\})
\end{equation}
\begin{equation}
\footnotesize
 h_v^{l} = COMBINE^{l}(h_v^{l-1}, a_v^{l} )
 \vspace{-0.2em}
\end{equation}
The \texttt{AGG} function collects information from the neighbors of $v$, $N(v)$, and extracts an embedding $a_v^{l}$ for the layer $l$. 
The \texttt{COMBINE} function updates the features of $v$ by combining $h_v^{l-1}$ with $a_v^{l}$. 
The updated embedding, $h_v^{l}$, captures information regarding $v$ and its neighborhood. 
After $L$ iterations of message passing, a \textit{read-out} is performed to generate a graph-level embedding, $h_\mathcal{G}$, which can be used for graph classification.

\section{Proposed MuxLink Attack Model}
\label{sec:Proposed_attack}
\textbf{Attack Model:} We assume an adversary located in the fab with access only to the GDSII representation of a locked design. 
The attacker performs reverse engineering to obtain the locked netlist and determines the location of the key-gates by tracing the key-inputs from the tamper-proof memory.
Fig.~\ref{fig:MuxLink} shows an overview of the main steps of MuxLink.

\vspace{-0.2em}
\subsection{Enclosing Subgraph Extraction}

The first step is identifying the key-controlled MUXes by tracing the key-inputs (see Fig.~\ref{fig:MuxLink}\raisebox{.5pt}{\textcircled{\raisebox{-.9pt} {1}}}) and removing them from the netlist. The netlist is then converted to an undirected graph $\mathcal{G}= (\mathcal{E},\mathcal{V})$, where $\mathcal{V}$ represents the set of nodes (gates), and $\mathcal{E} \subseteq \mathcal{V} \times \mathcal{V}$ represents the set of observed links (wires).
$\mA$ is the symmetric adjacency matrix of $\mathcal{G}$. 
The graph representation of the netlist does not include primary inputs and primary outputs, as we are interested in capturing the composition of gates and their connectivity. 
All the inputs to the MUXes are marked as target links, added to set $\mathcal{S}$ and excluded from $\mathcal{E}$  (see Fig.~\ref{fig:MuxLink}\raisebox{.5pt}{\textcircled{\raisebox{-.9pt} {2}}}). 
Next, MuxLink extracts an $h$-hop enclosing subgraph for each pair of target nodes $f$ and $g$ (see Fig.~\ref{fig:MuxLink}\raisebox{.5pt}{\textcircled{\raisebox{-.9pt} {3}}}). The $h$-hop enclosing subgraph for $(f,g)$ is induced from $\mathcal{G}$ containing the nodes $\{~j~ |~ d(j,f) \leq h ~\text{or}~ d(j,g)\leq h ~\}$, where $d(y,x)$ is the shortest path distance between $x$ and $y$. As discussed in Sec.~\ref{sec:backgroud_RW}, the techniques only check if the inputs to the MUXes $\{f_i,f_j\}$ are driving a single gate or multiple gates. However, the locking techniques do not consider the structure of the fan-out and fan-in cones of the selected gates, and hence, $\{f_i, f_j, g_i, g_j\}$ all have unique surroundings. Consequently, the extracted subgraphs in Fig.~\ref{fig:MuxLink}\raisebox{.5pt}{\textcircled{\raisebox{-.9pt} {3}}} are different and will have distinct link representations, allowing MuxLink to decipher the correct connections.

\vspace{-0.2em}
\subsection{Node Information Matrix Construction}
\label{sec:node_label}

A node information matrix $\mX$ is constructed for each extracted subgraph, where each node is associated with an $8$-bit one-hot encoded vector that encodes its Boolean functionality. E.g., the feature vector of node $g$ in Fig.~\ref{fig:MuxLink}\raisebox{.5pt}{\textcircled{\raisebox{-.9pt} {4}}} indicates that it is an XOR gate. For link prediction, the GNN must distinguish the target link and capture the relationship between the target nodes (colored in yellow in Fig.~\ref{fig:MuxLink}\raisebox{.5pt}{\textcircled{\raisebox{-.9pt} {4}}}) and the surrounding circuitry. 
To achieve this, we employ the double radius node labeling (DRNL)~\cite{SEAL}. 
Each node in the subgraph is assigned a tag that captures its relationship with the target link. 
Let $f$ and $g$ be the target nodes, the DRNL label $f_l(j)$ of a node $j$ is:

\begin{align}
   \vspace{-0.2em}
\footnotesize
f_l(j) = 1 + \text{min}(d_f, d_g) + (d / 2)[(d / 2) + (d \% 2) - 1]
   \vspace{-0.2em}
\label{hashing}
\end{align}
where $d_f = d(j,f)$, $d_g = d(j,g)$, and $d = d_f + d_g$. $(d / 2)$ is the integer quotient and $(d \% 2)$ is the remainder of $d$ divided by $2$. See Fig.~\ref{fig:MuxLink}\raisebox{.5pt}{\textcircled{\raisebox{-.9pt} {4}}} for an example of DRNL labeling. If $j$ has a path to only one of the target nodes, then $f_l(j)=0$. 
The target nodes are tagged with $1$ allowing the GNN to differentiate them from the rest of the gates. 
Each node's label is one hot-encoded and concatenated to its corresponding row in $\mX$. 
The dimension of $\mX$ depends on the largest assigned label in a given dataset, which depends on the target circuit and the subgraph size.
\vspace{-0.2em}
\subsection{Dataset Generation}
\label{sec:dataset}

MuxLink takes the graph representation of the target netlist $\mathcal{G}$ and extracts enclosing subgraphs for a set of sampled positive links (observed wires) and a set of sampled negative links (unobserved wires) for training. We generate a balanced dataset and use a maximum of $100,000$ training links. $10\%$ of the sampled links are kept for validation. The links between the target nodes are always removed from the subgraphs. During the attack phase, the enclosing subgraphs around the links in $\mathcal{S}$ are fed to the trained GNN, as shown in Fig.~\ref{fig:MuxLink}\raisebox{.5pt}{\textcircled{\raisebox{-.9pt} {5}}}. The GNN reports the likelihood score for each link.
\vspace{-0.2em}
\subsection{GNN Learning}
We employ the deep graph convolutional neural network (DGCNN)~\cite{zhang2018end} for graph classification. A DGCNN layer performs the following operation, where $\mH^{l+1}\in\mathbb{R} ^{n \times c_{l+1}}$ is the output embedding matrix of layer $l$, $c$ is the number of output channels, and $n$ is the number of nodes in the subgraph.
\begin{align}
   \vspace{-0.5em}
\footnotesize
 \mH^{l+1}=\sigma(\tilde{\mD}^{-1}(\mA +\mI)\mH^{l}\mB^{l})
    \vspace{-0.2em}
\end{align}
$\tilde{\mD}$ is the diagonal degree matrix, $\mB^{l}$ is a trainable weight matrix, and $\sigma(.)$ is a non-linear activation function. The initial embedding matrix is the same as the node information matrix $\footnotesize\mH^{0}=\mX$. After $L$ aggregation layers, the following concatenation is performed by DGCNN $\mH^{1:L}:=[\mH^1, ..., \mH^L]$ to represent a subgraph by a single vector. The tensor is then sorted row-wise according to $\mH^{L}$ and reshaped to $k(\sum_{l=1}^{L} c_l) \times 1$, selecting $k$ nodes to represent the subgraph. In MuxLink, we set $k$ such that $60\%$ of subgraphs have nodes less than or equal to $k$. Then, the final obtained embedding is fed to $1$-D convolutional layers for classification.

\vspace{-0.2em}
\subsection{Post-processing}
\label{sec:post_process}
The likelihood scores are processed to recover the secret key. Key prediction depends on the structure of the obfuscated locality and a controlled threshold parameter $th$. We describe the MuxLink post-processing approach for the different localities in the following subsections. 
\subsubsection{$S_{1}$ and $S_{5}$} Two key-inputs $\{k_{i}, k_{j}\}$ control two MUXes with the same inputs $\{f_i,f_j\}$. These strategies obfuscate two output nodes $\{g_i, g_j\}$. Hence, four links are considered during post-processing $\{(f_i,g_i),(f_j,g_i),(f_j,g_j),(f_i,g_j)\}$ denoted as $\{g_i1, g_i2,g_j1, g_j2\}$, respectively. The post-processing of such locked localities is outlined in Algorithm~1. $l_x\in [0, 1]$ is the likelihood score for link $x$. First, the absolute difference $\delta$ between the likelihood scores of the possible links for each gate $\{g_i, g_j\}$ is computed (lines 3-4). If none of the $\{\delta_1,\delta_2\}$ values is greater than $th$, MuxLink does not make a decision and reports $X$ for both key-bits (lines 18-19). Else, MuxLink checks which difference is larger ($\delta_1$ or $\delta_2$), then selects the link that has the highest likelihood score as the true link and predicts the key-bit value that passes that link. In the example shown in Fig.~\ref{fig:MuxLink}\raisebox{.5pt}{\textcircled{\raisebox{-.9pt} {6}}}, $\delta_1=|1-0.8|=0.2$ and $\delta_2=|0.9-0.4|=0.5$. If $th=0.01$, MuxLink will execute lines 11-17 as $\delta_2>\delta_1$. Since $l_{g_j1}>l_{g_j2}$, $k_j=1$ and $k_i=0$ (lines 12-13). 

\begin{algorithm}[tp]
\footnotesize
\begin{algorithmic}[1]
\STATE \textbf{Input:}  Threshold $th$, Likelihoods $L$\\
\STATE \textbf{Output:}  Deciphered keys $\{k_{i}, k_{j}\}$ \\
\STATE {$\delta_1=|l_{g_{i}1}-l_{g_{i}2}|$} \CommentSty{\footnotesize\blue{$//$Diff in likelihood scores for $g_i$}}
\STATE {$\delta_2=|l_{g_{j}1}-l_{g_{j}2}|$} \CommentSty{\footnotesize\blue{$//$Diff in likelihood scores for $g_j$}}
\IF{$\delta_1 >=th || \delta_2 >=th$} 
\IF{$\delta_1>\delta_2$}
\IF{$l_{g_{i}1}> l_{g_{i}2}$} 
\STATE{$k_i=0$, $k_j=1$}
\ELSE
\STATE{$k_i=1$, $k_j=0$}
\ENDIF
\ELSIF{$\delta_2>\delta_1$}
\IF{$l_{g_{j}1}> l_{g_{j}2}$}
\STATE{$k_i=0$, $k_j=1$}
\ELSE
\STATE{$k_i=1$, $k_j=0$}
\ENDIF
\ELSE
\STATE{$k_i=X$, $k_j=X$}
\ENDIF
\ELSE
\STATE{$k_i=X$, $k_j=X$}
\ENDIF
\RETURN{$\{k_{i}, k_{j}\}$}
\end{algorithmic}
\caption{\footnotesize\textsc{MuxLink post-processing for $\{S_1, S_4, S_5\}$}}\label{a1}
\end{algorithm}

\subsubsection{$S_{2}$ and $S_{3}$} Here, a single key-input $k_{i}$ controls one MUX, which obfuscates a single output node $g_i$. Hence, two links are considered during post-processing $\{g_i1, g_i2\}$. MuxLink computes the difference in likelihood scores between the two possible links, as follows: $\delta=|l_{g_{i}1}-l_{g_{i}2}|$. If $\delta<th$, MuxLink assigns $k_i=X$. However, if $\delta>=th$ and $l_{g_{i}1}>l_{g_{i}2}$, then $k_i=0$. If $\delta>=th $ and $ l_{g_{i}2}>l_{g_{i}1}$, then $k_i=1$.

\subsubsection{$S_{4}$} One key-input $k_{i}$ controls two MUXes, with the same inputs $\{f_i,f_j\}$ but in a different order. This strategy obfuscates two output nodes $\{g_i, g_j\}$. The same analysis outlined in Algorithm~1 is followed, but only $k_i$ is returned, ignoring $k_j$.

\section{Evaluation of Our MuxLink Attack Model}
\label{sec:results}

\begin{figure}[t]
\centering
\includegraphics[width=0.95\textwidth]{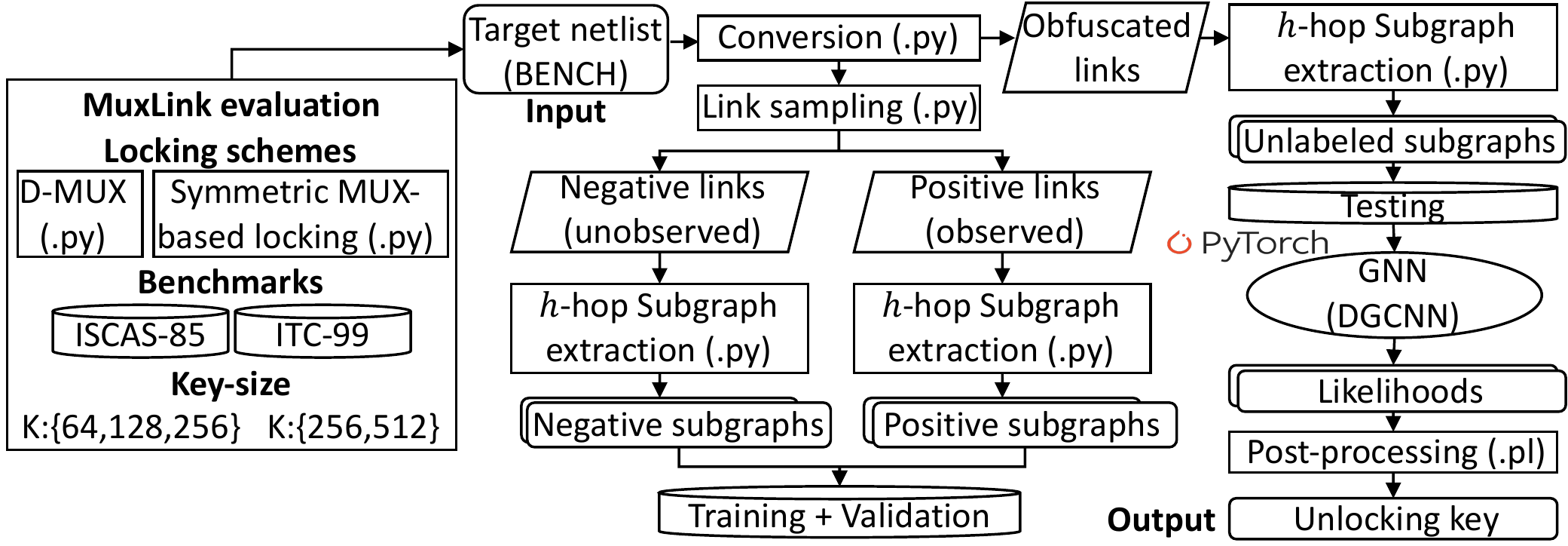}
\smallerspacecaption
\caption{Experimental setup and tool-flow.}
\label{fig:exp}
\end{figure}

We summarize the experimental setup and the process of dataset generation in Fig.~\ref{fig:exp}. 
We evaluate MuxLink on selected designs from the ISCAS-85 and ITC-99 combinational benchmarks locked using D-MUX~\cite{sisejkovic2021deceptive} and symmetric MUX-based locking~\cite{alaql2021scope}. 
We implement both techniques in Python as described in~\cite{sisejkovic2021deceptive,alaql2021scope}. 
We use the eD-MUX implementation of D-MUX.\footnote{To verify our implementation of the locking techniques, we launched SWEEP~\cite{sweep} and SCOPE~\cite{alaql2021scope} attacks on the locked benchmarks and observed the same attack resilience promised by the original work~\cite{sisejkovic2021deceptive,alaql2021scope} (see Fig.~\ref{fig:motivation}).} 
We implement the post-processing in Perl.

We lock the ISCAS-85 benchmarks with $K:\{64, 128, 256\}$, except for the \texttt{c1355} benchmark where $K=256$ was not achievable due to the small size of the design. 
Additionally, we lock the larger ITC-99 benchmarks with $K:\{256,512\}$, resulting in a total of $64$ locked designs. 
The benchmarks are locked and attacked in \textit{BENCH} format, following the methodology widely used by the LL community, and as also adopted by SWEEP and SCOPE attacks~\cite{sweep,alaql2021scope}.

\textbf{GNN Topology:} We start with the default DGCNN architecture of~\cite{zhang2018end}, which has four graph convolution layers with $\{32, 32, 32,1\}$ output channels, two 1-D convolutional layers with $\{16,32\}$ output channels, a fully-connected layer of $128$ neurons, a dropout layer with a dropout rate of $0.5$, and a \textit{softmax} layer of $2$ output units for classification. We use the \textit{tanh} activation function in the graph convolution layers and the \textit{ReLU} function in the rest of the layers. 
We use stochastic gradient descent with the Adam updating rule and train DGCNN for $100$ epochs with an initial learning rate of $0.0001$, and save the model with the best performance on the $10\%$ validation set to predict the testing links. MuxLink runs on a single machine utilizing $10$ cores ($2$x Intel(R) Xeon(R) CPU E5-2680 v4@2.4GHz) and a single NVIDIA-V100 GPU.

\textbf{Evaluation Metrics:} 
We use \blue{four} metrics for attack evaluation: \textit{accuracy} (AC), \textit{precision} (PC), KPA, \blue{and Hamming distance (HD)}. AC measures the ratio of correctly deciphered key-bits out of the entire key, i.e., $(K_{correct}/K_{total})\cdot{100\%}$. 
PC measures the ratio of correctly deciphered keys, counting every $X$ value as a correct guess, i.e., $((K_{correct} + K_{X})/K_{total})\cdot{100\%}$.
Finally, KPA measures the percentage of correctly deciphered key-bits out of the entire predictions, i.e., $(K_{correct}/(K_{total}-K_{X}))\cdot{100\%}$.

\begin{figure*}[!t]
\centering
\includegraphics[width=0.95\textwidth]{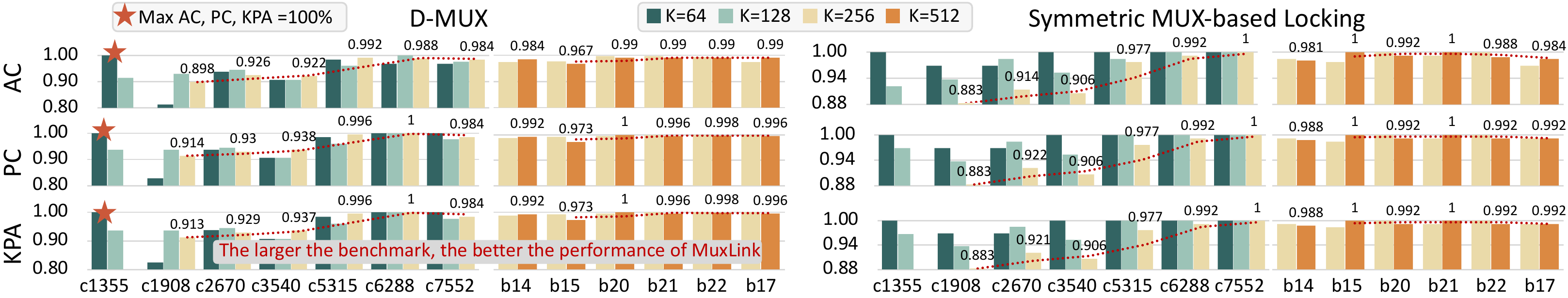}
\caption{Accuracy (AC), precision (PC), and KPA for MuxLink on the learning-resilient MUX-based locking techniques. 
The data labels are added for the ISCAS-85 benchmarks with $K=256$ and the locked ITC-99 benchmarks with $K=512$. 
We work on combinational counterparts of ITC-99 benchmarks.}
\label{fig:results}
\end{figure*}

\textbf{MuxLink Performance:} The AC, PC, and KPA of MuxLink on D-MUX and symmetric MUX-based locked benchmarks are presented in Fig~\ref{fig:results}, having $h=3$ and $th=0.01$.\footnote{Later we study the effect of $h$ and $th$ on the performance of the attack.} 
MuxLink achieves an average AC, PC, and KPA of $94.61\%$, $95.41\%$ and $95.37\%$, respectively, on the ISCAS-85 benchmarks locked using D-MUX, and an average AC, PC, and KPA of $98.49\%$, $99.43\%$ and $99.43\%$, respectively, on the ITC-99 benchmarks locked using D-MUX.
Additionally, MuxLink achieves an average AC, PC, and KPA of $96.95\%$, $97.31\%$ and $97.30\%$, respectively, on the ISCAS-85 benchmarks locked using symmetric MUX-based LL, and an average AC, PC, and KPA of $98.90\%$, $99.38\%$ and $99.38\%$, respectively, on the ITC-99 benchmarks locked using symmetric MUX-based LL. 
Overall, MuxLink achieves an AC, PC, and KPA up to $100\%$. 
These results show that MuxLink is capable of breaking the previously thought of as ``learning-resilient'' schemes.

\textbf{Effect of the LL Scheme:} The resilience of symmetric MUX-based LL against MuxLink is lower compared to D-MUX. 
Under the same $K$, symmetric MUX-based LL locks a fewer number of localities. 
This is because each obfuscated locality is controlled with two key-inputs $\{k_{i}, k_{j}\}$, with only two possible combinations $\{0, 1\}$ and $\{1, 0\}$. 
However, the equivalent implementation of $S_{4}$ in D-MUX (in addition to $S_{2}$ and $S_{3}$) is controlled via a single key-input. Therefore, D-MUX achieves a larger obfuscation under the same $K$.

\textbf{Effect of the Benchmark Size:}
The broken red lines in Fig.~\ref{fig:results} show the moving average of the score (AC, PC, and KPA) of MuxLink for the ISCAS-85 benchmarks locked with $K=256$ and for the ITC-99 benchmarks locked with $K=512$, versus the benchmarks (ordered from smallest to largest). 
The trend lines show that the performance of MuxLink enhances with the increase in the benchmark size. The larger the design is, the lower the impact of the obfuscation is. For instance, the KPA of MuxLink for the D-MUX locked \texttt{c1908} benchmark with $K=256$ is $92.27\%$, while the KPA for the D-MUX locked \texttt{c7552} benchmark with the same $K$ is $98.44\%$.
The plots also illustrate that, in general, the performance of MuxLink is better on the ITC-99 benchmarks compared to the performance on the smaller ISCAS-85 benchmarks.

\textbf{Effect of the Key-size ($K$):} The increase of $K$ faintly affects the performance of MuxLink on D-MUX. 
For instance, MuxLink KPA drops from $93.75\%$ to $92.94\%$ on the D-MUX locked \texttt{c2670}, when $K$ is increased from $K=64$ to $K=256$, respectively. 
Nevertheless, the average AC, PC, and KPA across the different key-sizes for the ISCAS-85 benchmarks are consistent around $95\%$ for the D-MUX scheme. 
However, the performance of MuxLink is slightly affected by the increase of $K$ when attacking symmetric MUX-based locking. 
For example, MuxLink KPA drops from $96.88\%$ to $92.13\%$ on the symmetric MUX-based locked \texttt{c2670}, when $K$ is increased from $K=64$ to $K=256$, respectively. 
On average, the KPA of MuxLink drops from $99.11\%$ to $94.65\%$, when moving from $K=64$ to $K=256$ on the symmetric MUX-based locked ISCAS-85 benchmarks.

\begin{figure}[t]
\centering
\includegraphics[width=0.85\textwidth]{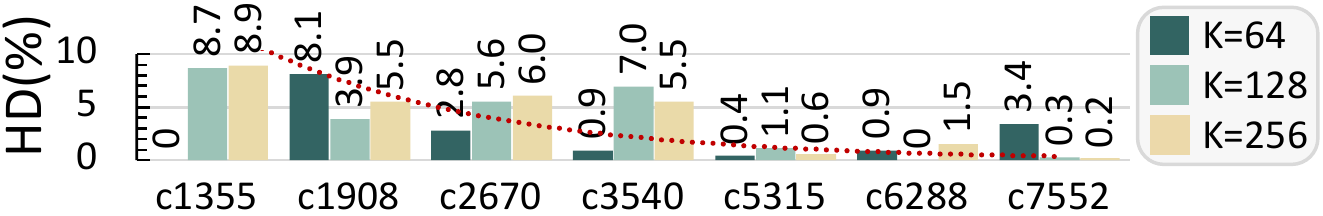}
\smallerspacecaption
\caption{Hamming distance (HD) between the outputs of original designs and the D-MUX locked designs recovered by MuxLink.}
\label{fig:HD}
\end{figure}

\textbf{Hamming Distance:} We compute the HD between the outputs of the recovered (D-MUX locked) design by MuxLink and those of the original design.
The goal of a defender is to enforce an HD of $50\%$ (maximum corruption), while the objective of an attacker is to recover the original design, i.e., obtain an HD of $0\%$. 
For each benchmark, we set the recovered key pattern by MuxLink and compute the HD by simulating $100,000$ random input patterns using \textit{Synopsys VCS}. 
For the cases where some key-bit values are undeciphered ($X$ values), we measure the HD for all the possible remaining key-bit assignments and compute the \blue{average}. 
We report the results in Fig.~\ref{fig:HD}.
The average HD value for the ISCAS-85 reconstructed by MuxLink is a mere $3.39\%$. 
Hence, using MuxLink, we (almost) determine the correct functionality.

\begin{figure}[t]
\centering
\includegraphics[width=0.85\textwidth]{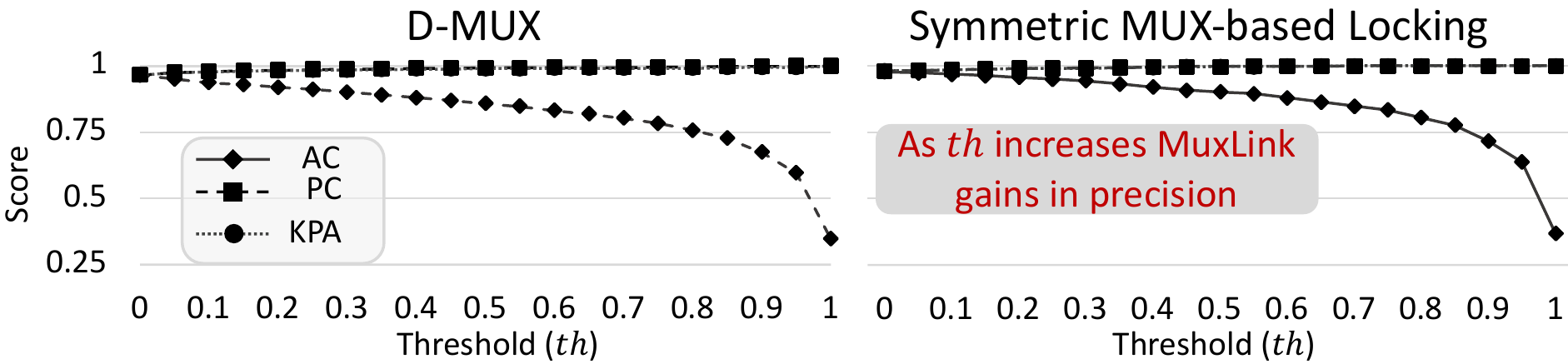}
\smallerspacecaption
\caption{MuxLink performance under different post-processing $th$ settings.}
\label{fig:th}
\end{figure}

\textbf{Post-processing Threshold ($th$):} We repeated the post-processing stage for a range of $th\in [0, 1]$ with a step of size $0.05$.
The GNN does not require any re-training as the $th$ value only affects the post-processing. This analysis is performed for both locking schemes. 
The average AC, PC, and KPA on the ISCAS-85 and ITC-99 benchmarks under the different $th$ settings are shown in Fig.~\ref{fig:th}. 
Setting a strict threshold of $th=1$ enforces a PC of $100\%$ for all the evaluated benchmarks. 
The ratio of the predicted key-bit values gets smaller with the increase in $th$ (reaches around $30\%$ for $th=1$). \blue{However, the small set of predicted keys is guaranteed to be correct.} Even with a $th=0$, MuxLink achieves an \blue{average} \blue{PC} of $96.54\%$ on the D-MUX locked benchmarks and $98.8\%$ on the symmetric MUX-based locked benchmarks.

\begin{figure}[t]
\centering
\includegraphics[width=0.85\textwidth]{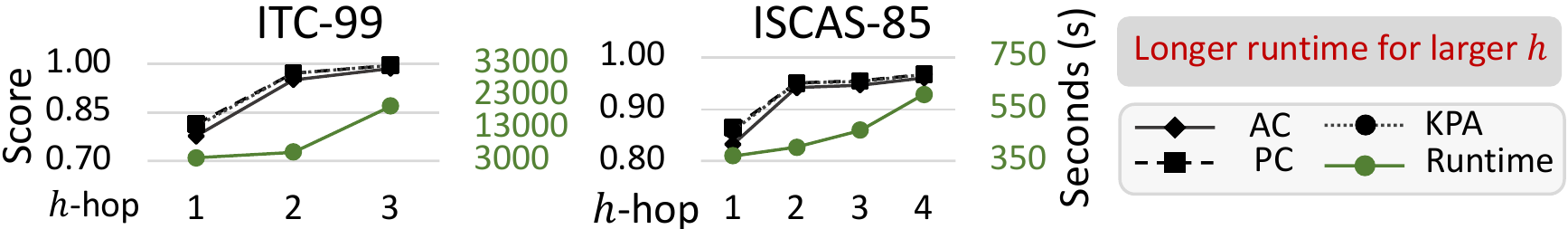}
\smallerspacecaption
\caption{MuxLink performance and runtime for different $h$-hop numbers.}
\label{fig:hop}
\end{figure}

\textbf{Subgraph Size and Runtime:} We study the effect of $h$-hop sampling on the performance and runtime of MuxLink.
We repeat the experiments with $th=0.01$ and vary $h\in[1,4]$ with a step size of $1$ (see Fig.~\ref{fig:hop}). 
The reported runtime includes subgraph sampling, training, testing, and post-processing. 
The performance of MuxLink in terms of AC, PC, and KPA improves with the increase in $h$ and saturates after $h\geq 3$. 
We primarily notice a jump in performance moving from $h=1$ to $h=2$. 
Nevertheless, the $1$-hop analysis sheds light on a fundamental vulnerability of the D-MUX and the symmetric MUX-based locking. 
Although the schemes claim protection at the locality level, MuxLink can decipher the obfuscated connections with high AC even when only considering the $1$-hop neighborhood of the obfuscated gates. 
With the increase in $h$, the number of neighbors and the runtime of MuxLink increase exponentially. Thus, we limit the hop size to $h=3$.

\textbf{Summary:} MuxLink was evaluated on two ``learning-resilient'' techniques: D-MUX~\cite{sisejkovic2021deceptive} and the symmetric MUX-based locking~\cite{alaql2021scope}. 
We consider different key-sizes, $h$-hop sizes, and threshold $th$ values. 
On average, MuxLink deciphers $96.87\%$ of the key-bits with a PC of $97.50\%$. The existing ML-based attacks~\cite{sail,snapshot,OMLA} fail to break these techniques because they try to extract non-existent key leakage.
However, MuxLink learns link formation and deciphers the key.

\section{Conclusion}
\label{sec:Conclusion}
We propose \textit{MuxLink} as a graph neural network (GNN)-based link prediction attack that successfully breaks the state-of-the-art learning-resilient D-MUX and symmetric MUX-locking, relying only on the structure of the locked design. 
The GNN learns the structure and connectivity of the target circuit around non-obfuscated wires, thereby generating meaningful link heuristics that help decipher the secret inputs to the locking MUXes. MuxLink achieves accuracy and precision up to $100\%$ on D-MUX and symmetric MUX-locked ISCAS-85 and ITC-99 benchmarks.
To the best of our knowledge, MuxLink is the first attack aimed at breaking learning-resilient logic locking. 
This work exposes a new source of exploitable leakage, i.e., link formation, and demonstrates that there is still a gap in designing learning-resilient logic locking.

\bibliography{main}
\bibliographystyle{IEEEtran} 

\end{document}